\documentclass[aps,prl,twocolumn,showpacs,superscriptaddress,groupedaddress]{revtex4}  
\usepackage{graphicx}  
\usepackage[caption=false]{subfig}
\usepackage{floatrow}
\usepackage{dcolumn}   
\usepackage{bm}        
\usepackage{amssymb}   
\usepackage{enumerate} 

\begin{document}


\title{Mass-density and Phonon-frequency Relaxation Dynamics of Under-coordinated Water Molecules}
\author{Chang Q Sun}
 \email{ecqsun@ntu.edu.sg}
\affiliation{Key Laboratory of Low-dimensional Materials and Application Technologies, and Faculty of Materials and Optoelectronics and Physics, Xiangtan University, Hunan 411105, China}
 \affiliation{School of Electrical and Electronic Engineering, Nanyang Technological University, Singapore 639798}

\author{Xi Zhang}
\affiliation{School of Electrical and Electronic Engineering, Nanyang Technological University, Singapore 639798}
\affiliation{ College of Materials Science and Engineering, China Jiliang University, Hangzhou 310018, China}

\author{Ji Zhou}
\affiliation{State Key Laboratory of New Ceramics and Fine Processing, Department of Materials Science and Engineering, Tsinghua University, Beijing 100084, China}

\author{Yongli Huang}
\affiliation{Key Laboratory of Low-Dimensional Materials and Application Technologies, and Faculty of Materials and Optoelectronics and Physics, Xiangtan University, Hunan 411105, China}

\author{Yichun Zhou}
\affiliation{Key Laboratory of Low-Dimensional Materials and Application Technologies, and Faculty of Materials and Optoelectronics and Physics, Xiangtan University, Hunan 411105, China}

\author{Weitao Zheng}
\email{wtzheng@jlu.edu.cn}
\affiliation{School of Materials Science, Jilin University, Changchun 130012, China}

\date{\today}

\begin{abstract}
The interplay between intra-molecular H-O covalent bond contraction, due to molecular under-coordination, and inter-molecular O:H expansion, due to inter-electron pair Coulomb repulsion, has been shown to be the source of the anomalous behavior of under-coordinated water molecules in nanoclusters and in the surfaces of water. The shortening of the H-O bond raises the local density of bonding electrons, which in turn polarizes the lone pairs of electrons on oxygen. The stiffening of the H-O bond increases the magnitude of O1s binding energy shift, causes the blueshift of the H-O phonon frequencies, and furthermore, elevates the melting point of molecular clusters and ultrathin films of water, which gives rise to their ice-like behavior at room temperature. At the same time, the elongation of the entire O:H-O bond polarizes and enlarges the under-coordinated H$_2$O molecules.

\end{abstract}

\pacs{61.20.Ja, 61.30.Hn, 68.08.Bc\\Supplementary Information is accompanied.}

\maketitle

\begin{figure}[!hbtp]
\includegraphics[width=2.8in,trim=150 230 250 160, clip]{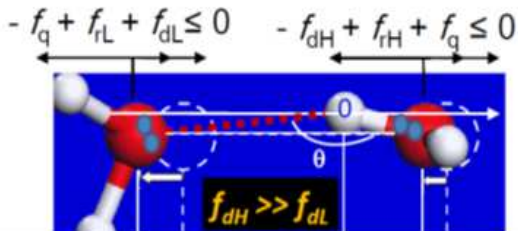}
\caption{Forces of inter-electron-pair (pairing dots) Coulomb repulsion $f_q$, restoring of deformation $f_{rx}$, and under-coordination-induced bond contraction $f_{dx}$ as well as the direction and degree of displacement for each O atom (in red) with respect to the H atom as the coordination origin. Subscript $L$ and $H$ represents the O:H and the H-O segment of the O:H-O hydrogen bond, respectively.}
\label{fig1}
\end{figure}

Under-coordinated water molecules refer to those with fewer than the ideal four neighbors as in the bulk of water \cite{kuhne13}. They occur in terminated hydrogen-bonded networks, in the skin of a large volume of water and in the gaseous state, and exhibit even more fascinating behavior than fully-coordinated ones \cite{liu96,ludwig01,michaelides07,turi05,verlet05,hammer04,gregory97,keutsch01,perez12}. For example, water droplets encapsulated in hydrophobic nanopores \citep{lakhanpal53,li12} and ultrathin water films on graphite, silica, and some metals \cite{michaelides07,xu10,miranda98,mcbride11,hodgson09,meng04,wang09} behave like ice at room temperature, i.e. such under-coordinated water molecules melt at a temperature higher than the melting point of water in bulk. (Empirically, the melting point is the temperature at which the vibration amplitude of an atom is increased abruptly to more than $3\%$ of its diameter \cite{omar93,lindemann10}.) More interestingly, the monolayer film of water is hydrophobic \cite{wang09,james11}. 
\\
\indent 
Molecular under-coordination enlarges the O1s core-level shift and causes a blue-shift of the H-O phonon frequency ($\omega_H$) of bulk water. The O1s level energy is 536.6 eV in the bulk of water \cite{winter07}, 538.1 eV in the surface of water and 539.8 eV in gaseous molecules \cite{abu09}. The $\omega_H$ phonon frequency has a peak centered at 3200 $cm^{-1}$ for the bulk, 3475 and 3450 $cm^{-1}$ for the surfaces of water and ice \cite{kahan07} and 3650 $cm^{-1}$ for gaseous molecules \cite{ceponkus12,shen06,buch04}. Such abnormal behaviors of electronic binding energy and phonon stiffness of under-coordinated water molecules are associated with a $5.9\%$ expansion of the surface O--O distance at room temperature \cite{liu96,abu08,wilson01,wilson02,lenz10}. In addition, the volume of water molecules confined to 5.1 and 2.8 nm TiO$_2$ pores increase by 4 and $7.5\%$, respectively, with respect to that in bulk \cite{solveyra13}. 
\\
\indent
Achieving a consistent understanding of these anomalies caused by molecular under-coordination remains a great challenge. In this article, we meet this challenge with a union of Goldschmidt, Feibelman and Paulings' (GFP) “under-coordination-induced atomic radius contraction” \cite{pauling47,goldschmidt27,feibelman96}, Anderson's ``strong localization'' \cite{abrahams79}, and our “O:H-O hydrogen bond” approach \cite{sun12}. Based on this framework, we show that under-coordination-induced GFP H-O bond contraction and the inter-electron-pair Coulomb repulsion dictate the observed attributes of enlarged O1s core-level and Raman frequency shifts, volume expansion, charge entrapment and polarization, as well as the “ice-like and hydrophobicity” nature of such water molecules at room temperature \cite{suppinfo}.
\\
\indent
Fig.\ref{fig1} illustrates the basic structural unit of the segmented ``O$^{\delta -}$:H$^{\delta +}$-O$^{\delta -}$" hydrogen bond (O:H-O will be used for simplicity) between under-coordinated water molecues \cite{sun12,suppinfo}. The ``:'' represents the electron lone pair of the sp$^3$-hybridized oxygen. $\delta$ is a fraction that denotes the polarity of the H-O polar-covalent bond and is determined by the difference in electronegativity of O and H. The hydrogen bond comprises both the O:H van der Waals (vdW) bond and the H-O polar-covalent bond, as opposed to either of them alone. The H-O bond is much shorter, stronger, and stiffer (0.1 nm, 4.0 eV, 3000 $cm^{-1}$) than the O:H bond (0.2 nm, 0.1 eV, 200 $cm^{-1}$)\cite{sunPRLrev}. The bond energy characterizes the bond strength while the vibration frequency characterizes the bond stiffness. In addition to the short-range interactions within the O:H and the H-O segments, Coulomb repulsion between the bonding electron pair ``-'' and the nonbonding electron lone pair ``:'' (the pair of dots on O in Fig.\ref{fig1}) is of vital importance to the relaxation of the O:H-O bond angle-length-stiffness under external stimulus \cite{sun12,sunPRLrev}. 
\\
\indent
In combination with the forces of Coulomb repulsion ($f_q$) and resistance to deformation ($f_{rx}$), each of the force $f_{dx}$ ($x = L$ represents the O:H and $x = H$ the H-O bond) can drive the hydrogen bond to relax. The two oxygen atoms involved in the O:H-O bond will move in the same direction simultaneously, but by different amounts with respect to the H atom that serves as the point of reference (Fig.\ref{fig1}). 

\vspace{-0.8em}

\vspace{-0.5em}
According to Goldschmidt \cite{goldschmidt27}, Feibelman \cite{feibelman96} and Pauling \cite{pauling47}, the radius of an atom shrinks once its coordination number (CN) is reduced. If the atomic CN is reduced relative to the standard of 12 in the bulk (for an $fcc$ structure) to 8, 6, 4, and 2, the radius will contract by 3, 4, 12, and 30$\%$, respectively \cite{goldschmidt27,feibelman96}. Furthermore, the bond contraction will be associated with a deepening of the inter-atomic potential well, or an increase of the bond energy \cite{sun07}, according to the general rule of energy minimization during the spontaneous process of relaxation. In other words, bonds between under-coordinated atoms become shorter and stronger. Such a bond order-length-strength (BOLS) correlation is formulated as follows \cite{sun07}: 

\vspace{-1.0em}
\begin{equation}
\label{eq1}
\left\{
\begin{array}{ll}
C_z  = \frac{d_z}{d_b} = 2 \left[1+ \exp(\frac{12-z}{8z}) \right]^{-1}   \\
C_z^{-m} =  \frac{E_z}{E_b}
\end{array}
\right.
\end{equation}

\begin{figure*}[!hbtp]
  \sidesubfloat[]{\includegraphics[width=2.1in, trim=35 25 60 80, clip]{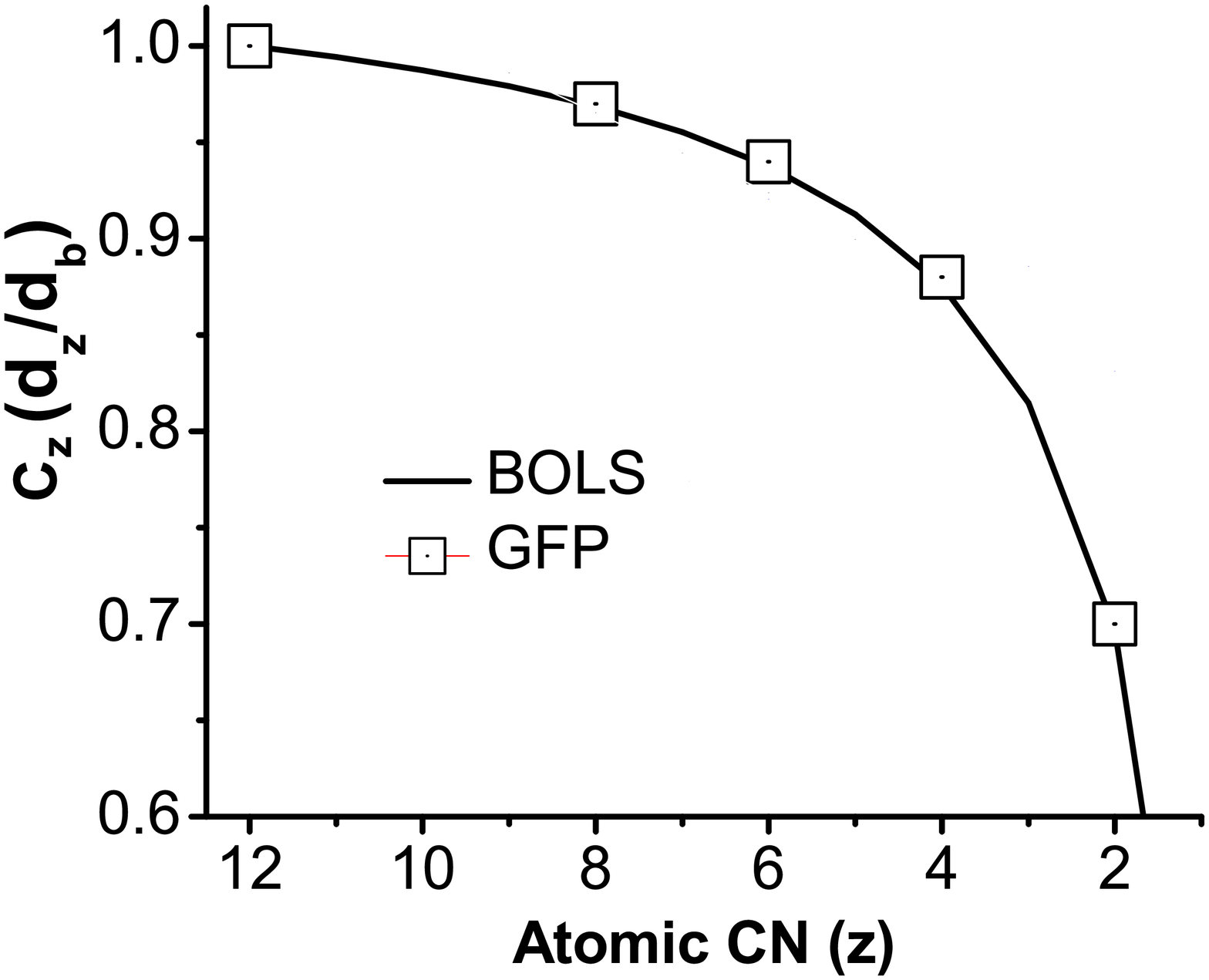}\label{fig2a}}
  \sidesubfloat[]{\includegraphics[width=2.1in, trim=35 25 120 50, clip]{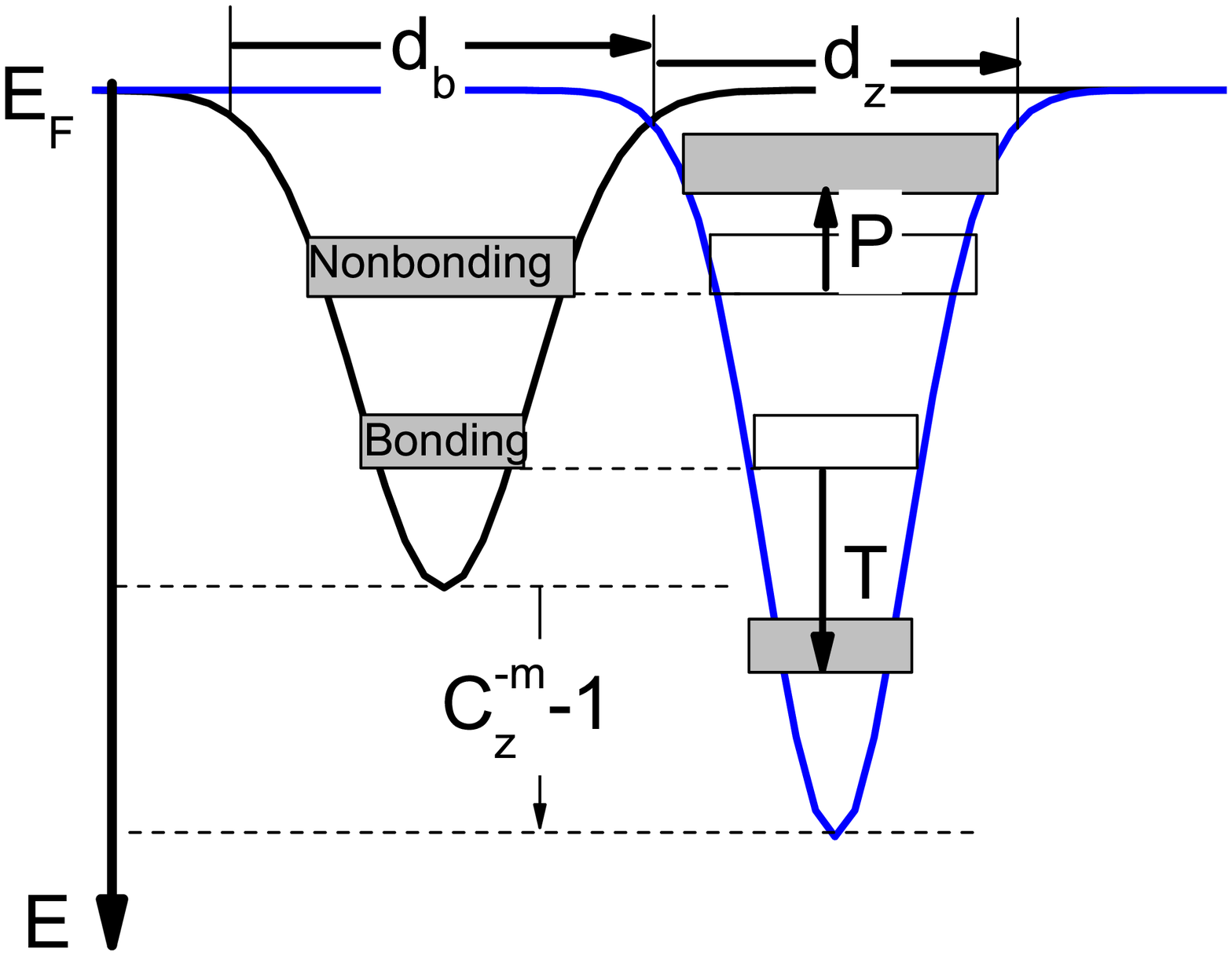}\label{fig2b}}
\vspace{-0.8em}
  \caption{BOLS corelation and nonbonding electron polarization (NEP, P) \cite{sun10,sun07}. (a) Bond order (CN) loss shortens and stengthens the bond, which causes core electron densification and entrapment ($T$) or potential well deepening. The binding energy of the core and bonding electrons will shift as the potential-well deepens. The densely entrapped core electrons in turn polarize the weakly bound nonbonding (lone pair) electrons ($P$), raising their energy closer to Fermi energy E$_f$.}
\label{fig2}
\end{figure*}

\vspace{-0.2em}
\noindent
where $m$ ($= 4$ for water \cite{zhao07}) relates the bond energy $E_z$ change with the bond length $d_z$. The subscript $z$ denotes the number of neighbors that an atom has, and $b$ denotes an atom in the bulk. Fig.\ref{fig2a} illustrates the coefficients of the bond contraction $(C_z - 1)$ \cite{sun07} and \ref{fig2b} the potential-well deepening $(C_z^{-m} - 1)$ due to bond contraction.

\floatsetup[figure]{style=plain,subcapbesideposition=top}

On the other hand, bond order (CN) loss causes a localization of electrons, according to Anderson \cite{abrahams79}. The bond contraction raises the local density of electrons in the core bands and electrons shared in the bonds. The electron binding energy in the core band will shift accordingly as the potential-well deepens (called entrapment, T). The densification and entrapment of the core and bonding electrons in turn polarize the nonbonding electrons (lone pair in this case), raising their energy closer to $E_F$, see Fig.\ref{fig2b} \cite{sun10}. 

However, molecular clusters, surface skins, and ultrathin films of water may not follow the BOLS precisely. An isolation of each H$_2$O molecule by the surrounding four lone pairs differentiates water’s response to the under-coordination effect from other materials. The binding energy difference between the O:H and H-O and the presence of the inter-electron-pair repulsion define the H-O covalent bond to be the ``master'' that contracts to a smaller degree than which Eq.\ref{eq1} predicts. The contraction of the H-O bond lengthens and softens the ``slave'' O:H bond by  the repulsion. 

Because of the difference in stiffness between the O:H and the H-O segments \cite{sun12}, the softer O:H segment always relaxes more in length than the stiffer H-O covalent bond does:  $|\Delta d_L|$$>$$|\Delta d_H|$. Meanwhile, the repulsion further polarizes the electron pairs during relaxation, which increases the viscosity of water. 

The relatively weaker O:H interaction contributes insignificantly to the Hamiltonian and its related properties, such as the core-level shift. However, the O:H bond length-stiffness relaxation determines the vibration frequency of the O:H phonons ($\omega_L$$<$300 cm$^{-1}$) \cite {sun12} and the energy for freezing a water molecule from a liquid state and the surface tension of liquid water \cite{zhao07}. 

The stiffening of the H-O bond increases the O1s core level shift, $\Delta E_{1s}$, elevates the critical temperature $T_C$ for phase transition, and increases the H-O phonon frequency $\omega_H$ according to the following relations \cite{sun04,sun05,sun06}:

\vspace{-1.0em}
\begin{equation}
\label{eq2}
\left.
\begin{array}{l}
T_c \\
\Delta E_{1s} \\
\Delta \omega_x
\end{array}
\right\}
\propto
\left\{
\begin{array}{l}
E_H\\
E_H\\
\frac{\sqrt{E_x}}{d_x} = \sqrt{Y_x d_x}
\end{array}
\right.
\end{equation}
	
\vspace{-0.2em}	
\noindent									
E$_x$ is the cohesive energy of the respective bond ($x=L$ or $H$). Theoretical reproduction of the critical temperature $T_C$ for ice VII-VIII phase transition under compression confirmed that the H-O bond energy determines the $T_C$ \cite{sun12}. The shift of the O1s binding energy from that of an isolated oxygen atom is also proportional to the H-O bond energy \cite{sun04}. Furthermore, the phonon frequency shift is proportional to the square root of the bond stiffness, which is the product of the Young's modulus ($Y_x \propto E_x d_x^{-3}$) and the bond length \cite{sun12,sunPRLrev}. 

The slight shortening of the H-O covalent bond and the significant lengthening of the O:H interaction result in the O:H-O bond elongation and molecular volume expansion. Further polarization and an increase in the elasticity and viscosity of the molecules will occur. For a molecular cluster of a given size, the BOLS effect becomes more significant as one moves away from the center. The smaller the molecular cluster, the stronger the BOLS effect will be, because of the higher fraction of under-coordinated molecules. Therefore, we expect that molecular clusters, ultrathin films, and the skin of the bulk of water could form an ice-like, low-density phase that is stiffer, hydrophobic, and thermally more stable compared to the bulk liquid.

In order to verify our hypotheses and predictions as discussed above, we calculated the angle-length-stiffness relaxation dynamics of the O:H-O bond and the total binding energy of water clusters as a function of the number of molecules $N$.  Structural optimizations of (H$_2$O)$_N$ clusters were performed using Perdew and Wangs' Dmol3 code (PW) \cite{perdew92} based on the general gradient approximation (GGA) and the dispersion-corrected OBS(PW) method \cite{ortmann06}, with the inclusion of hydrogen bonding and vdW interactions. The all-electron method was used to approximate the wave functions with a double numeric and polarization basis sets. The self-consistency threshold of total energy was set at $10^{-6}$ Hartree. In the structural optimization, the tolerance limits for the energy, force, and displacement were set at $10^{-5}$ Hartree, 0.002 Hartree/$\AA$ and $0.005 \AA$, respectively. Harmonic vibrational frequencies were computed by diagonalizing the mass-weighted Hessian matrix \cite{wilson80}. 

\begin{figure}[!hbtp]
\includegraphics[width=2.8in, trim=0 0 0 15, clip]{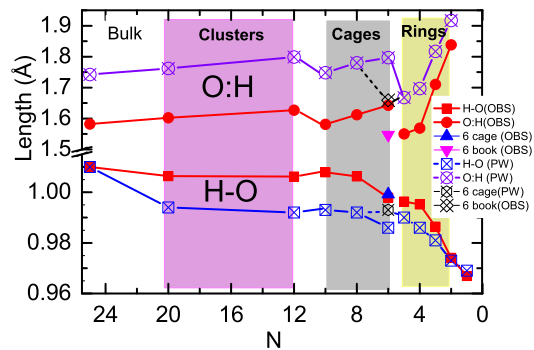}
\vspace{-0.6em}
\caption{Cluster size dependence of the O:H-O segmental lengths in the (H$_2$O)$_N$ clusters. The bond length was optimized using the PW \cite{perdew92} and the OBS(PW) \cite{ortmann06} methods. $N=6$ gives the three``cage'', ``book'', and ``prism'' hexamer structures, all with nearly identical binding energy \cite{perez12}.}

\label{fig3}
\end{figure}

Fig.\ref{fig3} shows the segment lengths of the O:H-O as a function of  (H$_2$O)$_N$ cluster size ($N$). Results optmized using the PW and the OBS algorithms exhibit the same trend of $N$-dependence. This comparison confirms that: 1) molecular CN-reduction shortens the H-O bond and lengthens the O:H, and 2) the shortening (lengthening) of the H-O bond is always coupled with the lengthening (shortening) of the O:H, independent of the algorithm.

As the $N$ is reduced from 24 (an approximation of the number of molecules in bulk water) to two (a dimer), the H-O bond contracts by $4\%$ from 0.101 nm to 0.097 nm and the O:H bond expands by $17\%$ from 0.158 to 0.185 nm, according to the OBS derivatives. This gives a $13\%$ expansion of O--O distance, which is a significant amount for the dimer. The O:H and H-O length profiles are non-monotonic because of different effective CNs in different structures \cite{suppinfo}. The monotonic O:H and H-O relaxation profiles for $N$$\leq$6 will be discussed in the subsequent sections without influencing the generality of conclusions.

\floatsetup[figure]{style=plain,subcapbesideposition=top}
\begin{figure}[!hbtp]
  \sidesubfloat[]{\includegraphics[width=1.6in, height=2.0in,  trim=0 0 0 20, clip]{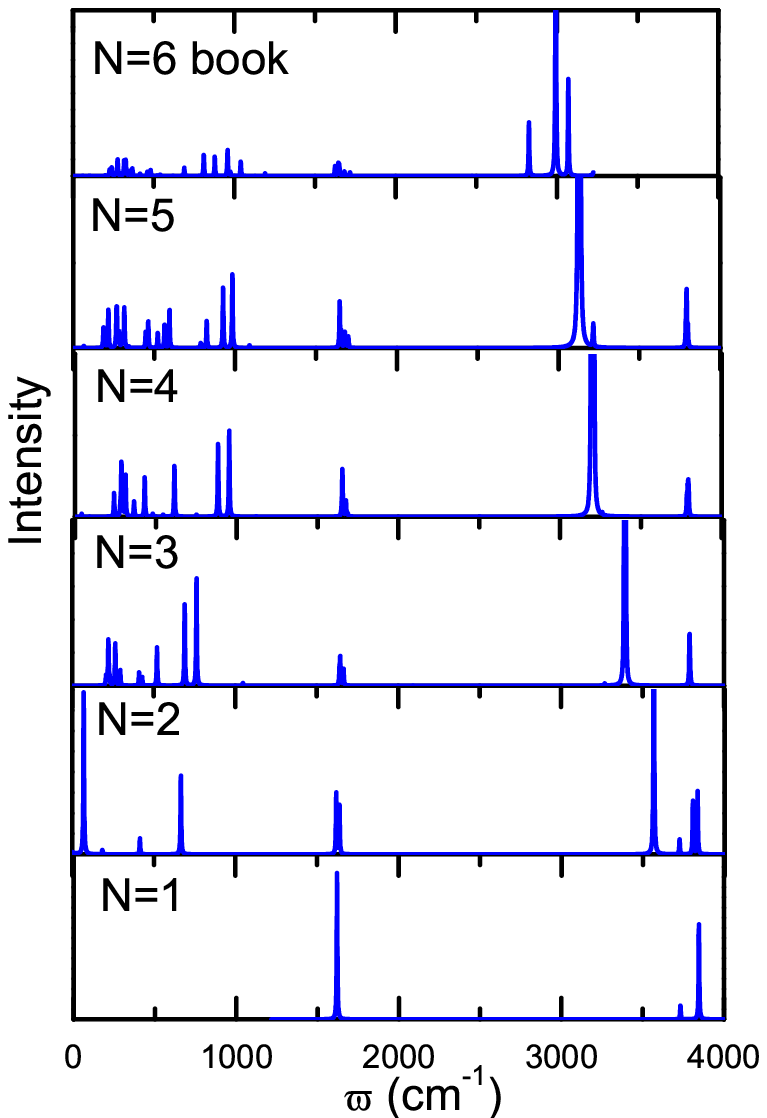}\label{fig4a}}\\
  \sidesubfloat[]{\includegraphics[width=3.0in, height=1.6in, keepaspectratio]{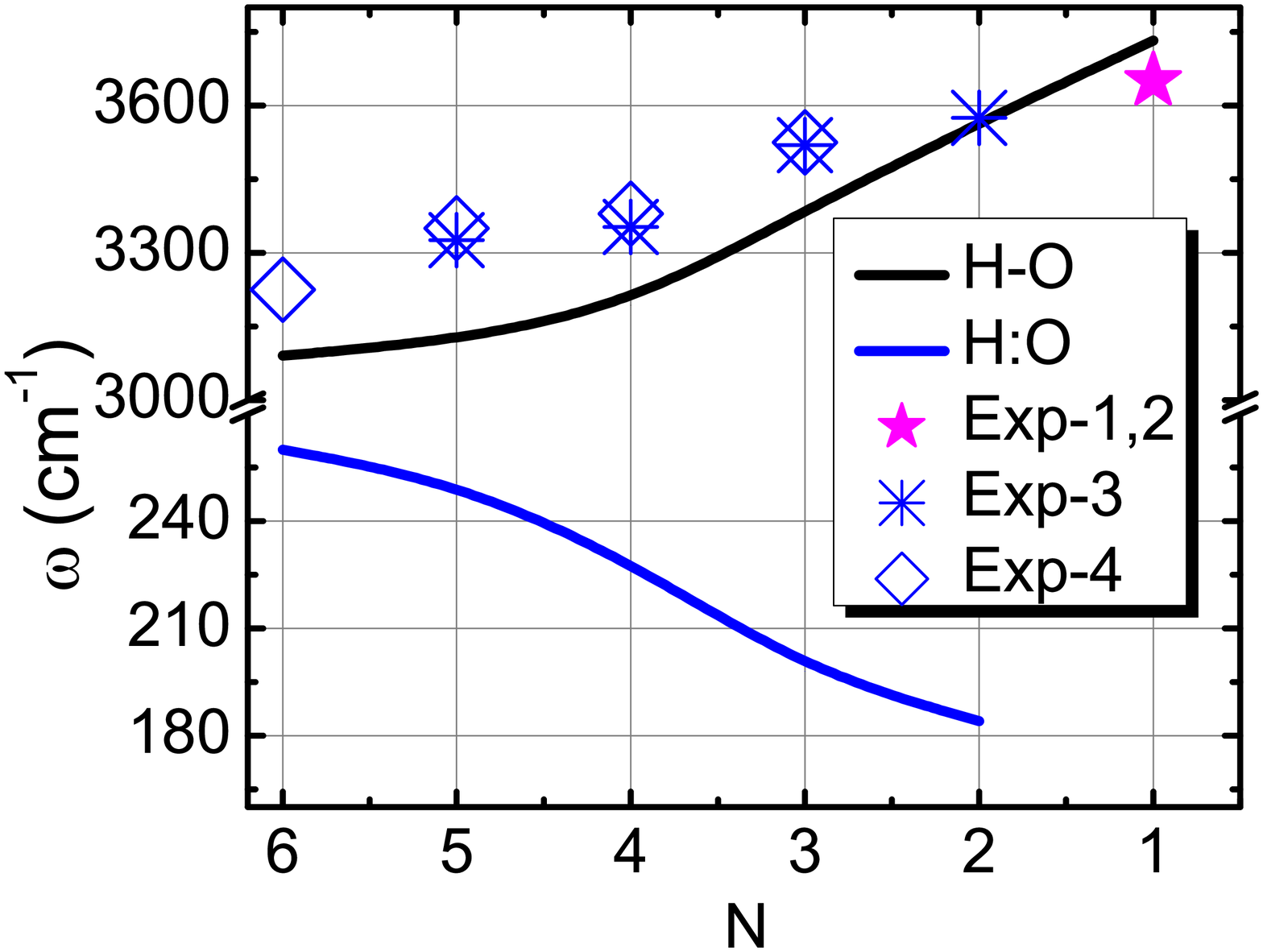}\label{fig4b}}
  \caption{(a) Size dependence of the phonon spectra of (H$_2$O)$_N$ clusters ($N \leq 6$). (b) The calculated (solid line) $\omega_H$ blue-shift has a similar trend as the measured frequencies (scattered data) of the H-O phonons of (H$_2$O)$_N$, shown as Exp-1 \cite{cross37}, Exp-2 \cite{qsun09}, Exp-3 \cite{ceponkus12}, and Exp-4 \cite{hirabayashi06}. Measurements of the $\omega_L$ redshift are not presently available due to experimental limitations.}
\label{fig4}
\end{figure}

\begin{figure}[!hbtp]
  \sidesubfloat[]{\includegraphics[width=3in, height=1.8in,  trim=0 0 0 20, clip, keepaspectratio]{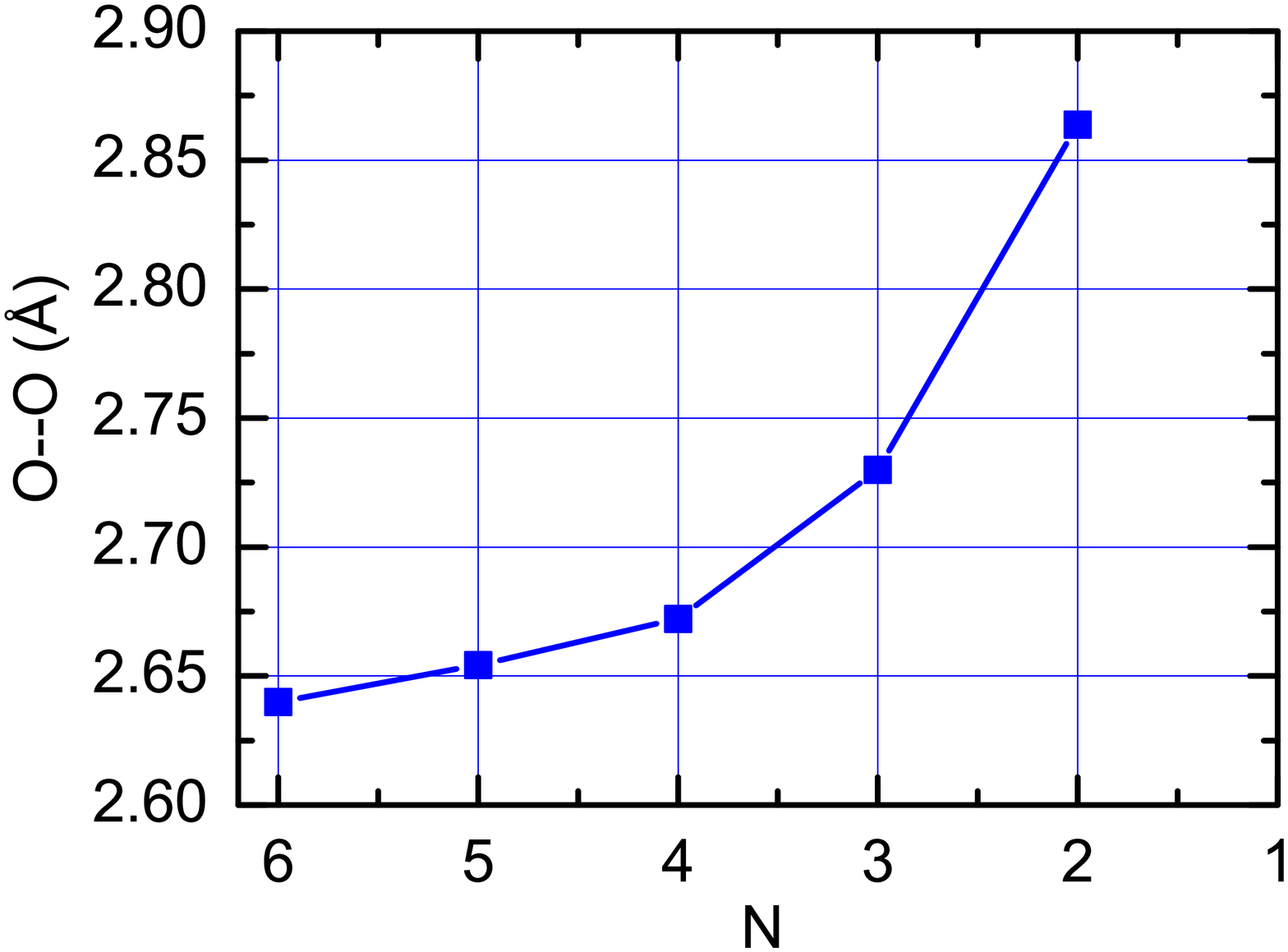}\label{fig5a}}\\
  \sidesubfloat[]{\includegraphics[width=3.3in, height=1.8in, keepaspectratio]{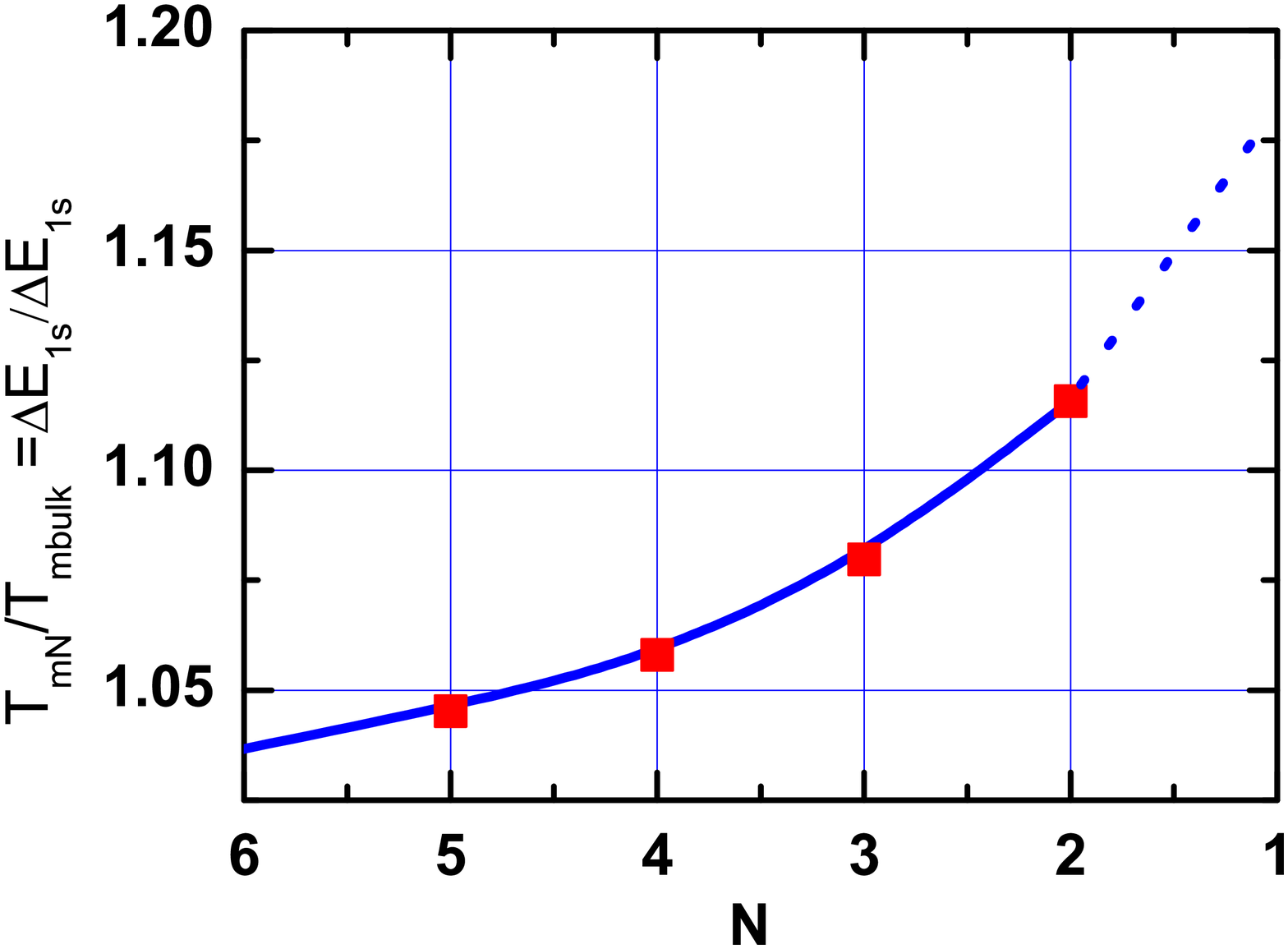}\label{fig5b}}
  \caption{$N$-dependence of (a) the O--O distance, (b) the melting point, $T_{mN}$, (to $N = 2$ for dimers) and the O1s core-level shift (to $N=1$ for gas monomers) of (H$_2$O)$_N$ clusters based on numerically derived values of $\left(\frac{d_{HN}}{d_{HB}}\right)$ and Eq.\ref{eq2}.}
\label{fig5}
\end{figure}

Fig.\ref{fig4a} shows the $N$ dependence of the (H$_2$O)$_N$ vibration spectra. As expected, the stiffer $\omega_H$ ($>2700$ $cm^{-1}$) experiences a blue-shift while the softer $\omega_L$ undergoes a red-shift as the $N$ is reduced. The $\omega_L$ shifts from 250 to 180 $cm^{-1}$ as the (H$_2$O)$_6$ becomes a dimer (H$_2$O)$_2$. The O:H-O bending mode (400-1000 $cm^{-1}$) shifts slightly but the H-O-H bending mode ($\approx$ 1600 $cm^{-1}$) remains the same. The calculated $\omega_H$ blue-shift in Fig.\ref{fig4b} agrees with that measured in molecular clusters \cite{ceponkus12,buch04,qsun09,hirabayashi06,pradzynski12} and in ice and water surfaces \cite{kahan07} (see Fig.S1 and S2 \cite{suppinfo}). This consistency validates our predictions regarding the under-coordination-induced asymmetric phonon relaxation dynamics of water molecules. 
\\
\indent
Fig.\ref{fig5a} plots the $N$-dependence of the O--O distance derived from Fig.\ref{fig3}. According to our calculations, the O--O distance expands by $8\%$, when the $N$ is reduced from 20 to 3, which is compatible to the value of $5.9\%$ measured in the water surface at $25^\circ C$ \cite{wilson02}. The polarization enhancement of the under-coordinated water molecules \cite{gregory97,yang10} is related to the O--O distance because of the charge conservation of the molecules. As it has been discovered using an ultra-fast liquid jet vacuum ultra-violet photoelectron spectroscopy \cite{siefermann10}, the dissociation energy for an electron in solution changes from a bulk value of 3.3 eV to 1.6 eV at the water surface. The dissociation energy, as a proxy of work function and surface polarization, decreases further with molecule cluster size (Fig.S3 in \cite{suppinfo}). These findings verify our predictions on the under-coordination-induced volume expansion and polarization of water molecules. \\
\indent
The polarization of molecules caused by both under-coordination and inter-electron-pair repulsion enhances the elasticity and the viscosity of the skin of water. The high elasticity and the high density of surface dipoles form the essential conditions for the hydrophobicity of a contacting interface \cite{sun09}. Therefore, given our established understanding of high elasticity and polarization in under-coordinated water molecules, it is now clear why the monolayer film of water is hydrophobic \cite{wang09}.

\floatsetup[figure]{style=plain,subcapbesideposition=top}

Fig.\ref{fig5b} shows the predicted $N$-dependence of the melting point ($T_{mN}$) elevation and the O1s energy shift ($\Delta E_{1sN}$). According to Eq.\ref{eq2}, both $T_{mN}$ and $\Delta E_{1sN}$ are proportional to the H-O bond energy in the form of: $\frac{T_{mN}}{T_{mB}}= \frac{\Delta E_{1sN}}{\Delta E_{1sB}} = \frac{E_{HN}}{E_{HB}}=\left(\frac{d_{HN}}{d_{HB}}\right)^{-4}$. Subscript $B$ denotes the bulk. One can derive from the plots that when the $N$ is reduced from a value of infinitely large to two, the $T_{mN}$ will increase by $12\%$ from 273 K to 305 K. It is now clear why the ultrathin water films \cite{michaelides07,xu10,miranda98,mcbride11,hodgson09,meng04,wang09} or water droplets encapsulated in hydrophobic nanopores \cite{lakhanpal53,li12} behave like ice at room temperature. The expected O1s energy shift ($C_z^{-4}-1$) of water clusters also agrees with the trend of the measurements (see Fig.S4 \cite{suppinfo}). For instance, the O1s core level shifts from 538.2 to 538.6 eV and to 539.8 eV, when the water cluster size is reduced from $N = 200$ to 40 and to free water molecules \cite{abu09,bjorneholm99}. 
\\
\indent
A hybridization of the GFP H-O bond contraction \cite{pauling47,goldschmidt27,feibelman96,sun07}, Anderson localization \cite{abrahams79,sun10}, and the segmented hydrogen bond premise \cite{sun12,sunPRLrev} has enabled clarification of the observed anomalous behavior of under-coordinated water molecules. Agreement between numerical calculations and experimental observations has verified our hypothesis and predictions:
\\
\indent
i) GFP contraction of the H-O bond and O:H elongation dictate the unusual behaviour of water molecules in the nanoscale O:H-O networks and in the skin of water.
\\
\indent
ii )The shortening of the H-O bond raises the density of the core and bonding electrons in the under-coordinated molecules, which in turn polarizes the nonbonding electron lone pairs on oxygen. 
\\
\indent
iii) The stiffening of the H-O bond increases the O1s core-level shift, causes the blue-shift of the H-O phonon frequency, and elavates the melting point of water molecular clusters, surface skins, and ultrathin films of water.
\\
\indent
iv) Under-coordinated water molecules could form an ice-like, low-density phase that is hydrophobic, stiffer, and thermally more stable than the bulk water \cite{liu96,ludwig01}.

Special thanks to Philip Ball, Yi Sun, Buddhudu Srinivasa and John Colligon for their comments and expertise. Financial support from NSF China (Nos.: 21273191, 1033003, 90922025) is gratefully acknowledged.

\end{document}